\documentstyle[preprint,aps]{revtex}

\begin{document}
\draft
\preprint{ }
\title {Center of mass and relative motion in time dependent density functional
theory}

\author{ G. Vignale}
\address{Institute of Theoretical Physics, University for California, Santa
Barbara,
California 93106-4030 \\  and \\  Department of Physics, Missouri University,
Columbia,  Missouri
65211} \date{\today}
\maketitle

\begin{abstract}
It is shown that the exchange-correlation part of the action functional
$A_{xc}[\rho (\vec r,t)]$ in
time-dependent density functional theory  ,  where $\rho (\vec r,t)$ is
the time-dependent density, is invariant under the transformation to an
accelerated frame of
reference $\rho (\vec r,t) \to \rho ' (\vec r,t) = \rho (\vec r + \vec x
(t),t)$,  where $\vec x (t)$
is an arbitrary function of time.  This invariance implies that the
exchange-correlation potential in
the Kohn-Sham equation transforms in the following manner: $V_{xc}[\rho '; \vec
r, t] =
V_{xc}[\rho; \vec r + \vec x (t),t]$. Some of the approximate formulas that
have been proposed for
$V_{xc}$ satisfy this exact transformation property, others do not.  Those
which  transform in the
correct manner automatically satisfy the ``harmonic potential theorem",  i.e.
the separation of the
center of mass motion for a system of interacting particles in the presence of
a harmonic external
potential.  A general method to generate functionals which  possess the correct
symmetry is proposed.

\end{abstract}

\pacs{  }

\narrowtext
Time dependent density functional theory (TDFT)\cite{Runge,Kohn,Gross} is a
very valuable tool for the
study of the dynamic behavior of correlated electronic systems under the
influence of external probes
which can be represented as  time-dependent potentials.  The range of problems
to which this
formalism is successfully applied is  expanding \cite{Gross}:  it includes the
calculation of atomic
collision cross sections,  the  behavior of atoms under very strong
electromagnetic fields,  the
calculation of  excitation energies in atoms, molecules, and solids .  Crucial
to the success of the
theory is the availability of simple approximations for  the
exchange-correlation potential $V_{xc}[\rho;\vec r , t]$,  which is an ordinary
 function  of $\vec
r$  and $t$,  but a complicated, non-local functional of the density
distribution $\rho (\vec r' ,
t')$ for  times $t'$ earlier than $t$.   Examples of  approximations for
$V_{xc}$ are the ``adiabatic
time dependent local density approximation" (ATLDA)  of Zangwill and Soven
\cite{Zangwill},  and the
local linear response approximation, with a frequency-dependent
exchange-correlation
kernel,  of Gross and Kohn \cite{Kohn}.   In constructing such approximations,
one should
try to make sure that the approximate form satisfies as many of the known exact
properties of the
functional as possible.  One example of
 rigorous constraints   on admissible approximate functionals is provided by
the scaling relations for
the static density functional theory \cite {Levy}.   Another type of
constraint, which is peculiar to
the time-dependent theory,  has been recently pointed out by  Dobson \cite
{Dobson}.  He
has observed that,  according to a general ``harmonic potential theorem"
(HPT),  a system of
interacting electrons confined by an external harmonic potential  $V_0 (\vec r)
= (1/2) \vec r \cdot
{\bf K} \cdot \vec r$,  must  possess solutions in which the dynamics of the
center of mass is
completely decoupled from that of the internal  degrees of freedom.
Assuming that the system is initially in the ground state,  the center of mass
wave function is a
minimum uncertainty gaussian centered about the origin.  When a  time-dependent
 uniform   electric
field $\vec E (t)$  is applied to such a system, the time dependence of the
center of mass wave
function is such that the  gaussian is rigidly transported (to within a
time-dependent phase
factor),   the position of its center being determined by the solution of the
classical equation of
motion
\begin {equation} m {d^2 \vec R_{CM} (t) \over dt^2} = - e \vec E
(t) - {\bf K} \cdot \vec R_{CM} (t) -{e \over c} {d \vec R_{CM} (t) \over
dt}\times \vec B , \label
{eq1} \end {equation}
 where $\vec B$ is a uniform and constant magnetic field.
 The fact that the
``relative" dynamics is not affected by the external force  implies that the
electronic density must
also be rigidly transported, following the motion of the center of mass, i.e.,

\begin {equation}
\rho (\vec r ,t) = \rho_0(\vec r - \vec R_{CM} (t))
\label {eq2}
\end {equation}
where $\rho_0 (\vec r)$ is the static electronic density  in the absence of the
driving field.
Of course,  one would hope that approximate versions of  TDFT were able to
reproduce this
elegant exact result.  Dobson's discovery \cite {Dobson}  is that this  is not
always  the
case.   For example, the Zangwill-Soven \cite {Zangwill} adiabatic
ATLDA does satisfy this requirement,  but  the Gross-Kohn\cite {Kohn}
linearized  response
theory, which  includes a frequency dependence  (i.e.,  a memory)   in the
exchange-correlation
kernel,  does not. Other approximations,  based on inhomogeneous hydrodynamics,
 also fail to
satisfy the HPT \cite {Dobson} . The root of the difficulty  was traced by
Dobson to the
inclusion of non-locality in time but not in space.  He  developed a simple
modification of the TDLDA,
whereby the density modulation induced by the driving field is decomposed in
two parts,  one
corresponding to a rigid shift of the whole distribution  (i.e,  a motion of
the center of mass),
the other  including the remaining internal deformations.  The Gross-Kohn
formalism is applied only
to the latter part,  while  the center of mass component is still treated in
the adiabatic LDA.
This  formalism satisfies the HPT.

	In this paper,  we demonstrate that the occurrence of the HPT in TDFT is in
fact the consequence
of a simple symmetry  of the exchange-correlation part of the action.  The
approximations that
Dobson finds to satisfy HPT are the ones that satisfy this symmetry,  and those
that violate the HPT
are the ones that violate this symmetry.  Furthermore,  the cure he proposes to
make the TDLDA
consistent with the HPT  can be viewed as an instance of a general procedure
whereby the correct
symmetry of the exchange-correlation potential can be built at the outset in
any approximation.

We now present our results in detail.  The basic theorem of TDFT  is the
Runge-Gross  theorem,
according to which,  for any  time-dependent density $\rho (\vec r , t)$
defined within an
appropriate class of functions in the interval $t_0<t<t_1$,  there exists an
essentially unique time
dependent potential $V(\vec r ,t)$, such that the solution $\vert \psi (t)
\rangle$ of the
time-dependent Schroedinger equation \begin {equation}
(i{ \partial \over \partial t } - \hat H_0 - \hat V (t)) \vert \psi (t) \rangle
 =0,
\label {eq3}
\end {equation}
with the condition that  $\psi(t_0)$ is the Hohenberg-Kohn ground-state wave
function corresponding to
the density $\rho_0(\vec r) = \rho(\vec r,t_0)$,  yields $\rho (\vec r ,t)$ as
the expectation value
$\langle \psi (t) \vert \hat \rho (\vec r) \vert \psi (t) \rangle$  of the
density operator $\hat
\rho (\vec r) = \sum_{i=1}^N \delta(\vec r  - \vec r_i)$.  Here \begin
{equation}
\hat H_0 = \sum_{i=1}^N {1 \over 2m} (\vec p_i  + {e \over c} \vec A (\vec
r_i))^2 +  {e^2 \over 2}
\sum_{i,j} {1 \over \vert \vec r_i  - \vec r_j \vert }
 \label {eq4}
\end {equation}
is the familiar Hamiltonian for N interacting electrons in the presence of a
uniform magnetic field
$\vec B$  described by the vector potential $\vec A (\vec r) = \vec B  \times
\vec r /2$,  and
\begin {equation}
\hat V  (t)  = \sum_{i=1}^N V(\vec r_i ,t)
\label {eq5}
\end {equation}
is the time-dependent potential.
 The  expression  ``essentially unique"  refers to the possibility of adding to
$V(\vec r
,t)$ an arbitrary function $f(t)$ of  time,  which causes the wave function to
be multiplied by
a phase factor $exp[i \int f(t) dt ]$  without affecting the density. The
``appropriate class" of
densities  includes all the densities for which the potential $V(\vec r,t)$
exists, and is Taylor
expandable,  with finite convergence radius, in a neighborhood of $t_0$. Under
the assumption that at
time $t_0$ the system is in the ground state - an assumption that we make in
the rest of this
paper -  the Runge-Gross theorem establishes the existence of an essentially
unique mapping from
time-dependent densities to time-dependent potentials.

Let us now look at the density $\rho (\vec r , t)$  from the point of view of
an  accelerated
observer whose position,  relative to the original reference frame,  is given
by the function  $\vec x
(t)$.  It is assumed that the accelerated observer uses cartesian axes that
remain parallel to the
axes of the original reference frame,  i.e.,  there is no rotation.  The
density seen by the
observer,  in terms of his own $\vec r$ coordinate is
$\rho ' (\vec r,t) = \rho (\vec r + \vec x (t),t)$. It is assumed that $\vec
x(t_0) = d \vec x(t)/dt
(t=t_0) = 0$ so that both the density and the wave function seen by the
accelerated observer coincide
with those seen by the inertial observer at time $t_0$.  Our first important
result is that there
exists  a  time-dependent potential  $V'(\vec r , t)$  that  generates $\rho '
(\vec r ,t)$  when the
Schroedinger equation  is solved.  The explicit form for $V'(\vec r,
t) $, aside from an  inessential additive function of time, is \begin
{equation} V' (\vec r ,
t)  =  V(\vec r + \vec x (t),t) + m \vec a(t) \cdot \sum_{i=1}^N \vec r_i  +
(e/c)  \vec (v(t) \times
\vec B) \cdot \sum_{i=1}^N \vec r_i , \label {eq6}
\end {equation}
  where $\vec a (t)$  and  $\vec v (t)$  are the
second and first derivatives of $\vec x (t)$  with respect to time.  The
wavefunction  $\psi
'$
corresponding to this transformed potential  is related to the  wavefunction
$\psi$
corresponding to potential $V$   by the transformation
\begin {equation}
\vert \psi ' (t) \rangle  =  \Pi_{i=1}^N \hat U_i (t) \vert \psi (t) \rangle
\label {eq7}
\end {equation}

where the unitary  transformation $\hat U_i (t)$  is defined as

\begin {equation}
\hat U_i (t) = exp[ -i \vec r_i \cdot (m \vec v(t) - {e \over 2c} \vec B \times
\vec x (t)) ] exp [i
\vec p_i \cdot \vec x (t)]. \label {eq8}
\end {equation}

In writing the above formulas we have assumed  that there is a uniform and
constant magnetic field
$\vec B$ acting on the electrons.  This is because we want to discuss later the
generalized harmonic potential theorem in the presence of a magnetic field.
The familiar situation
is recovered by putting $B=0$  in the above formulas.  The detailed
mathematical proof of the key
equations ~(\ref{eq6}) and ~(\ref{eq7}) is provided in the appendix.  However,
the correctness
of these equations is physically evident.  Equation ~(\ref{eq6})  says that the
 potential seen  by the accelerated observer  is the original potential
expressed in terms of
the new coordinate,  plus an inertial force $-m \vec a (t)$ that couples like a
uniform gravitational
field to the center of mass of the system,  plus  a uniform electric  field
$\vec v \times \vec B /c$
that arises from the transformation of the uniform magnetic field.
Equation ~(\ref{eq8})  says that the new wavefunction is obtained from the
former  by translating the
coordinate of each electron  by  $\vec x (t)$  and the momentum of each
electron   by  $m \vec v
(t)$.  When the magnetic field is present an additional translation  of the
momentum  by  $- e \vec
B \times \vec x (t) /2c  $  is required in order to restore the original gauge
for the vector
potential. Equations ~(\ref{eq6}) and ~(\ref{eq7})   enable us to determine the
transformation of the
``internal" action functional,  defined  as follows:
\begin {equation}
\tilde A [\rho] = \int_{t_0}^{t_1} \langle \psi (t) \vert i {\partial \over
\partial t} - \hat H_0
\vert \psi (t) \rangle  dt
\label {eq9}
\end {equation}
(Note the absence of the external potential in this definition).
We obtain
\begin {equation}
\tilde A[\rho '] =  \tilde A[\rho] +N \int_{t_0}^{t_1} [m \vec a(t) \cdot \vec
R_{CM}(t)+{e \over
c} (\vec v(t) \times \vec B) \cdot \vec R_{CM} (t)+ {mv^2(t) \over 2}+{e \over
2c}(\vec B \times \vec
x (t)) \cdot \vec v (t)]dt,
\label {eq10}
\end {equation}
where
\begin {equation}
N \vec R_{CM} (t) = \int \vec r \rho (\vec r, t)  d\vec r
\label{eq101}
\end{equation}
  is the coordinate of the center of
mass of wave function $\psi$.
The key observation is that,  because the additional forces in the
accelerated frame of reference depend only on the electron coordinates,  the
additional terms in the
transformed action  can be expressed explicitely in terms of the density.
Clearly,  these additional
terms would have had exactly the same form  if we had considered the
transformation of the
non-interacting action functional
\begin {equation}
\tilde A_0 [\rho] = \int_{t_0}^{t_1} \langle \psi_0 (t) \vert i {\partial \over
\partial t} - \hat T
\vert \psi_0 (t) \rangle  dt
\label {eq11}
\end {equation}
where $\psi_0$  is the   wave function corresponding to the density $\rho (\vec
r ,t)$
in a {\it non-interacting} system,  and $\hat T$ is the kinetic energy
operator.

Now,  according to Runge and Gross the interacting and non-interacting action
functionals are
related as
\begin {equation}
\tilde A [\rho] = \tilde  A_0 [\rho] +  A_H [\rho]+A_{xc} [\rho]
\label {eq12}
\end {equation}
which constitutes a definition of the ``exchange-correlation"  part of the
action functional.  Since
the Hartree part of the functional
\begin {equation}
 A_H [\rho] =  - {e^2 \over 2} \int_{t_0}^{t_1} dt \int d \vec r \int d \vec r'
{\rho (\vec r,t) \rho
(\vec r',t) \over \vert \vec r - \vec r' \vert}
 \label {eq13}
\end {equation}
is manifestly invariant under the transformation $\rho \to \rho
'$  we conclude,  comparing  the transformation equations for $\tilde A$ and
$\tilde A_0$  that
$A_{xc}$  is  invariant  under this transformation:
\begin {equation}
A_{xc} [\rho '] = A_{xc}[\rho]
\label {eq14}
\end {equation}

This is the most important result in this paper.  The tranformation law for the
exchange correlation
potential is derived in the following way.  We consider the difference  between
the $xc$ action
calculated for two neighboring densities  $\rho$   and $\rho + \delta \rho$:
\begin {equation}
 A_{xc} [\rho + \delta \rho] - A_{xc} [\rho ]  = \int_{t_0}^{t_1}dt'  \int d
\vec r'  V_{xc}[\rho;
\vec r' ,t'] \delta \rho (\vec r ', t')
\label {eq15}
\end {equation}
 which follows from  the definition of $V_{xc}$  as a first functional
derivative  of $A_{xc}$  with respect to the density.  We can write the same
relation for the
transformed densities  $\rho '$   and  $\rho ' + \delta \rho '$:
\begin {equation}
 A_{xc} [\rho' + \delta \rho'] - A_{xc} [\rho' ]  = \int_{t_0}^{t_1}dt'  \int d
\vec r'   V_{xc}[\rho'; \vec r' ,t'] \delta
\rho' (\vec r ', t')
\label {eq16}
\end {equation}
  Now,  according to  eq. ~(\ref{eq14})   the
left hand sides of these two equations must be equal.  It follows then that the
right hand sides are
also equal.  By doing a change of variable in one of the two integrals and
using the fact that
$\delta \rho$  is arbitrary,  we easily establish the  transformation  property
\begin {equation}
V_{xc}[\rho ';\vec r ,t] = V_{xc} [\rho; \vec r +\vec x (t),t]
\label {eq17}
\end {equation}

We now show that  an exchange-correlation potential that correctly transforms
according to
equation  ~(\ref{eq17})  automatically satisfies the ``harmonic potential
theorem". We suppose that
for $t \leq t_0$  the density  of the interacting electron system is described
by the usual  Kohn-Sham
equation  with a {\it time-independent} external {\it harmonic} potential
$V_0(\vec r)$,  and a
static Hartree potential $V_H(\rho_0;\vec r)$ and exchange correlation
potential
$V_{xc}[\rho_0 ;\vec r]$.  $\rho_0 (\vec r)$  is the unperturbed ground-state
density,  given by
the sum of the squares of the N lowest lying eigenfunctions  of the Kohn-Sham
equation
\begin {equation} [ \hat T +  V_0(\vec r) + V_H[\rho_0; \vec r] +
V_{xc}[\rho_0; \vec r]] \phi_i(\vec r) =\epsilon_i \phi_i(\vec r).
\label {eq179}
\end {equation}
The time dependence of these orbitals is given by $\phi_i (\vec r,t) =
\phi_i(\vec r) exp [-i
\epsilon_i t]$,  and they satisfy the time-dependent equation
\begin {equation} [i {\partial \over
\partial t} - \hat T -  V_0(\vec r) - V_H[\rho_0; \vec r] - V_{xc}[\rho_0; \vec
r]] \phi_i(\vec r,t)
=0. \label {eq18}
\end {equation}
  The system is now perturbed by a  uniform  time-dependent
electric field $\vec E (t)$:  the  center of mass moves according to eq.
{}~(\ref{eq1})  and the harmonic potential
theorem says that the density must evolve according to  eq. ~(\ref{eq2}).  Our
task is to show that
in fact  $\rho(\vec r, t) = \rho_0 (\vec r - \vec R_{CM} (t))$  is a
self-consistent solution of the
time-dependent Kohn-Sham equation  {\it  in the presence of the driving
electric field}:
\begin {equation}
[i {\partial \over \partial t} - \hat T -  V_0(\vec r) - V_H[\rho; \vec r ,t] -
V_{xc}[\rho;
\vec r,t] - e \vec E(t) \cdot \vec r] \phi^E_i(\vec r,t) = 0
\label {eq19}
\end {equation}
To prove this fact,  we observe that the Kohn-Sham   equation, in the presence
of the
driving field, eq. ~(\ref{eq19}),  can be generated starting from the Kohn-Sham
 equation without the
driving field, eq. ~(\ref{eq18}),  simply by applying to the latter a
transformation to  an
accelerated frame of reference with  $\vec x(t) = -\vec R_{CM}  (t)$ where
$\vec R_{CM} (t)$ is
the solution of the equation of motion ~(\ref{eq1}).   In fact,  using eq.
{}~(\ref{eqA6}) of the
Appendix, it is easy to prove that the transformed  wave functions  $\phi_i '
(\vec r ,t) =  \hat U
\phi_i  (\vec r ,t)$  satisfy the equation
\begin {eqnarray}
& & [i {\partial \over \partial t} - \hat T -  V_0(\vec r - R_{CM}(t)) -
V_H[\rho_0; \vec r -
R_{CM}(t)] - V_{xc}[\rho_0; \vec r - \vec R_{CM}(t) ]
\nonumber \\
& & +m \vec a_{CM}(t) \cdot \vec r +{e \over
c} (\vec v_{CM}(t) \times \vec B)\cdot \vec r + {mv_{CM}^2(t) \over 2}+{e \over
2c}(\vec B \times
\vec  R_{CM} (t)) \cdot \vec v_{CM} (t)]]\phi'_i(\vec r,t) = 0
\label{eq20}
\end{eqnarray}

Now,  we substitute the harmonic potential form for
\begin {equation}
V_0(\vec r - \vec R_{CM}) = V_0 (\vec r) - \vec R_{CM} \cdot {\bf K} \cdot \vec
r + {1 \over 2}
\vec R_{CM} \cdot {\bf K} \cdot \vec R_{CM},
\label {eq201}
\end{equation}
and  we use the  transformation of the Hartree and exchange-correlation
potentials
$V_{xc}[\rho_0;\vec r - \vec R_{CM}) = V_{xc}[\rho; \vec r]$.  Using  the
equation of motion
essentially equivalent to the Kohn-Sham equation ~(\ref{eq19}) in the presence
of the driving
field.   The  only difference is  a time dependent additive term in the
potential  which can be
eliminated by further multiplying  each wavefunction by the phase factor $exp
[iS(t)]$   where
\begin {equation}
 S(t) = \int_{t_0}^{t_1} [{m \over 2} v_{CM}^2(t) - {1 \over 2} \vec R_{CM}(t)
\cdot {\bf K} \cdot \vec R_{CM} (t) +{e \over c} \vec A(\vec R_{CM}(t))\cdot
\vec v_{CM}(t)]dt
\label {eq21}
\end {equation}
  is the classical action
for the motion of the center of mass.  Since the new orbitals $\phi_i'  $ (with
the additional
phase factor incorporated) yield the density   $\rho_0 (\vec r - \vec R_{CM}
(t))$  and satisfy the
time-dependent Kohn-Sham equation, in the presence of the driving field,  with
 Hartree and
exchange-correlation potentials evaluated at that same density,  we conclude
that they {\it are}  the
self-consistent solution of the driven  Kohn Sham equation.  Any approximation
for $V_{xc}$  that
satisfies eq.~  (\ref{eq17})  will automatically satisfy the HPT.

Consider the adiabatic  local density approximation \cite {Zangwill}.  The
action functional in this
approximation  has the form
\begin {equation}
A_{xc}[\rho (\vec r,t)]  =  -\int_{t_0}^{t_1} dt\int d \vec r\epsilon_{xc}(\rho
(\vec r ,t))
\label {eq22}
\end {equation}
where $\epsilon_{xc}(\rho)$ is the exchange-correlation energy {\it density}
of the uniform electron
gas of density $\rho$.   This
expression  is manifestly invariant under the transformation  $\rho \to \rho'$
since the latter
amounts to a simple change of variable in the space part of the integral.
Therefore,  eq.

On the other hand,  consider the linear response theory of Gross and Kohn \cite
{Kohn} for $V_{xc}$.
It has the   form
\begin {equation}
V_{xc}[\rho; \vec r,t] =  V_{xc}^{LDA}[\rho_0 (\vec r); \vec r] +
\int_{t_0}^{t_1} f_{xc}[\rho_0
(\vec r);t-t'] \delta \rho (\vec r ,t') dt'
\label {eq23}
\end {equation}
where it is assumed that the density $\rho (\vec r,t) = \rho_0(\vec r)+ \delta
\rho (\vec r,t)$
deviates only slightly from the static equilibrium density.  $V_{xc}^{LDA}
[\rho_0(\vec r);\vec r]$
is the usual local density approximation for the static density $\rho_0$.
Notice that the kernel
$f_{xc}$  is a function of a time difference,  but only one position   (i.e. we
have locality in
space but not in time).  In the case of harmonically confined electrons we know
from eq. ~(\ref{eq2})
that $\delta \rho (\vec r,t) = -\vec R_{CM} (t) \cdot \vec \nabla \rho_0(\vec
r)$ if $\vec R_{CM}
(t)$ is small. On the other hand,  from  eq. ~(\ref{eq17}), we  know that
$V_{xc}[\rho; \vec r,t] =
V_{xc}[\rho_0, \vec r - \vec R_{CM}(t)] \sim  V_{xc}[\rho_0, \vec r] - \vec
R_{CM}(t) \cdot \vec
\nabla V_{xc}[\rho_0;\vec r]$, and $V_{xc}[\rho_0;\vec r]
=V_{xc}^{LDA}[\rho_0;\vec r]$ .
Substituing this  in  eq. ~(\ref{eq23}) we see that  the integral on the right
hand side must equal
$\vec \nabla V_{xc}^{LDA}[\rho_0,\vec r] \cdot \vec R_{CM}(t)$.  Because this
must be true for an
arbitrary (small) driven motion of $\vec R_{CM}(t)$   we  see that the only
admissible
time dependence of $f_{xc}(t-t')$ is proportional to a $\delta $  function of
$t-t'$, i.e.  the only
admissible approximation in this class must be local in time as well as in
space.  Since the
Gross-Kohn  potential is non-local in time,  it violates eq. ~ (\ref{eq17}  )
and therefore also the
harmonic potential theorem.

How can one make sure that approximate forms of the exchange-correlation
potential satisfy
eq.~(\ref{eq17} )?  A  general way is to start from an action that depends only
on the ``relative
density"  defined as follows.  For a  given density  $\rho$  we construct the
position of the center
of mass,  and then  we refer $\rho $  to an accelerated frame in which the
center of mass is at
rest.  The  so defined density
\begin {equation}
\rho_{rel}(\vec r , t) =  \rho (\vec r + \vec R_{CM} (t),t)
\label {eq24}
\end {equation}
 is what we call the relative
density.  By construction,  it is invariant with respect to transformation to
an arbitrary
accelerated frame of reference.  Therefore,  if the action functional  is
written as a functional
of the relative density $A_{xc}[\rho (\vec r, t] = \bar A_{xc}[\rho_{rel} (\vec
r, t)]$ it will
automatically satisfy  the  symmetry embodied by eq.~ (\ref{eq14}). The
exchange-correlation potential
will then be given by
\begin {equation}
V_{xc}[\rho; \vec r, t] = {\delta \bar A_{xc}[\rho_{rel}]  \over  \delta
\rho_{rel}} (\vec r - \vec
R_{CM} (t) ,t) .
\label {eq25}
\end {equation}
 An example of application of this method is the procedure proposed by
Dobson\cite {Dobson} to go
beyond the adiabatic linearized LDA without violating the HPT.  The idea is to
divide the
density variation into two parts:  one corresponding to a rigid shift of the
center of mass, the
other including the remaining internal deformations.
The Gross-Kohn  frequency dependent kernel  is applied only to the relative
density variation,
while the center of mass variation is still treated in the ordinary ATLDA.

 In the language of this paper, this idea corresponds to the following
procedure. Write the relative
density as $\rho_{rel} (\vec r,t) = \rho_0(\vec r) + \delta \rho_{rel}
(\vec r ,t)$,  where $\rho_0(\vec r)$  coincides with the absolute initial
density, because   the
center of mass is initially at the origin of the absolute coordinates.
 Of course, the correction $\delta \rho_{rel}$ vanishes in the case of
uniformly driven harmonically
confined system,  but it need not vanish in more general cases  (for example,
in a   non-uniformly
driven harmonic system).  Approximate the action functional as
  \begin {eqnarray}
& & \bar A_{xc} [\rho_{rel} (\vec r,t)] = -\int_{t_0}^{t_1} dt\int d \vec
r\epsilon_{xc}(\rho_0
(\vec r)) \nonumber \\ & &+{1 \over 2} \int_{t_0}^{t_1} dt \int_{t_0}^{t_1} dt'
\int d\vec r
f_{xc}[\rho_0(\vec r);t-t'] \delta \rho_{rel} (\vec r ,t)  \delta \rho_{rel}
(\vec r ,t'), \label
{eq26} \end {eqnarray}
where $f_{xc}[\rho_0, t-t']$ is the Gross-Kohn exchange-correlation kernel.
The
exchange-correlation potential is then constructed by taking the functional
derivative of $\bar
a_{xc}$ with respec to $\rho_{rel}(\vec r,t)$, and translating the position
vector from $\vec r$
to $\vec r - \vec R_{CM}(t)$,  as indicated by eq. ~(\ref {eq25}).  $\vec
R_{CM}(t)$ itself  must be
determined self-consistently  from the knowledge of the full density,
according to eq.

We note,  in closing, that all our result could be easily generalized to the
case of a uniform
time-dependent magnetic field.  The HPT generalizes to this  case.  The only
difference is the
appearance of the additional  electric field  $(e/2c) d \vec B(t)/dt \times
\vec x(t)$    in eq. ~
(\ref{eq6}).

\medskip
{\bf Acknowledgements}
I acknowledge support from NSF Grant No. DMR~9100988 and hospitality
at the ITP where part of this work was done, under NSF Grant No. PHY89-04035.

{\bf Appendix - Proof of eqs. (6)  and (7)}

We start from the Schroedinger equation eq.~(\ref{eq3})  for  $\psi$ and apply
to it the unitary
transformation  $\hat U (t)$  defined in eq.
in the following manner
\begin {equation}
\hat U (t)(i{ \partial \over \partial t } - \hat H_0 - \hat V (t))  \hat
U^{-1}(t) \hat U(t) \vert
\psi (t) \rangle  =0
\label{eqA1}
\end {equation}
We then observe that
\begin {eqnarray}
\hat U(t)   i{\partial \over \partial t} \hat U^{-1}(t) &=& i {\partial \over
\partial t} - [m \vec
a(t) +{e \over 2c} (\vec v(t) \times \vec B)] \cdot  \sum_{i=1}^N \vec r_i
\nonumber  \\ &+&\sum_{i=1}^N \vec p_i \cdot
\vec v(t) + N m v^2 (t) - {e \over 2c} (\vec v(t) \times \vec B) \cdot \vec
x(t),
\label {eqA2}
\end{eqnarray}

and
\begin {equation}
\hat U(t)  \vec r_i \hat U^{-1}(t) =  \vec r_i +\vec x(t)
\label {eqA3}
\end{equation}
\begin {equation}
\hat U(t)   \vec p_i \hat U^{-1} (t)= \vec p_i + m \vec v(t) -{e \over 2c}\vec
B \times \vec x (t)
\label {eqA4}
\end{equation}

The density associated with the wavefunction  $\psi' = \hat U (t) \psi$
clearly is $\rho '$.
Furthermore
\begin {equation}
\hat U(t)  [\hat H_0 + \hat V] \hat U^{-1}(t) = \hat H_0 + \sum_{i=1}^N V(\vec
r_i +
\vec x(t),t) + N {m v^2 (t) \over 2} + \sum_{i=1}^N \vec p_i \cdot \vec v(t)
+{e \over 2c}
\sum_{i=1}^N \vec r_i \cdot (\vec v (t) \times \vec B)
\label {eqA5}
\end {equation}

Therefore,  the transformed Schroedinger equation  has the form
\begin {eqnarray}
& &[i {\partial \over \partial t} - \hat H_0 - \sum_{i=1}^N V(\vec r_i + \vec
x(t),t) -m \vec a(t)
\cdot \sum_{i=1}^N \vec r_i
\nonumber  \\
& & -{e \over c} \sum_{i=1}^N \vec r_i \cdot (\vec v (t) \times \vec B) +
N {m v^2 \over 2} +{e \over 2c} (\vec B \times \vec x(t)) \cdot \vec v(t)]
\vert \psi' (t) \rangle =
0
\label {eqA6}
\end {eqnarray}

This can be seen as the time evolution of the wavefunction in the presence of a
transformed
time-dependent potential  given by eq. ~(\ref{eq6}  ) apart from an additive
time dependent
constant.   This completes the proof of eqs. (6) and (7).

\end{document}